\documentclass[12pt]{article}
\usepackage{html}
\usepackage{epsf}
\usepackage{graphicx}
\usepackage{amssymb}
\begin{document}
\title{MATTERS OF GRAVITY, The newsletter of the APS Topical Group on 
Gravitation}
\begin{center}
{ \Large {\bf MATTERS OF GRAVITY}}\\ 
\bigskip
\hrule
\medskip
{The newsletter of the Topical Group on Gravitation of the American Physical 
Society}\\
\medskip
{\bf Number 40 \hfill Fall 2012}
\end{center}
\begin{flushleft}
\tableofcontents
\vfill\eject
\section*{\noindent  Editor\hfill}
David Garfinkle\\
\smallskip
Department of Physics
Oakland University
Rochester, MI 48309\\
Phone: (248) 370-3411\\
Internet: 
\htmladdnormallink{\protect {\tt{garfinkl-at-oakland.edu}}}
{mailto:garfinkl@oakland.edu}\\
WWW: \htmladdnormallink
{\protect {\tt{http://www.oakland.edu/?id=10223\&sid=249\#garfinkle}}}
{http://www.oakland.edu/?id=10223&sid=249\#garfinkle}\\

\section*{\noindent  Associate Editor\hfill}
Greg Comer\\
\smallskip
Department of Physics and Center for Fluids at All Scales,\\
St. Louis University,
St. Louis, MO 63103\\
Phone: (314) 977-8432\\
Internet:
\htmladdnormallink{\protect {\tt{comergl-at-slu.edu}}}
{mailto:comergl@slu.edu}\\
WWW: \htmladdnormallink{\protect {\tt{http://www.slu.edu/colleges/AS/physics/profs/comer.html}}}
{http://www.slu.edu//colleges/AS/physics/profs/comer.html}\\
\bigskip
\hfill ISSN: 1527-3431

\bigskip

DISCLAIMER: The opinions expressed in the articles of this newsletter represent
the views of the authors and are not necessarily the views of APS.
The articles in this newsletter are not peer reviewed.

\begin{rawhtml}
<P>
<BR><HR><P>
\end{rawhtml}
\end{flushleft}
\pagebreak
\section*{Editorial}

The next newsletter is due February 1st.  This and all subsequent
issues will be available on the web at
\htmladdnormallink 
{\protect {\tt {https://files.oakland.edu/users/garfinkl/web/mog/}}}
{https://files.oakland.edu/users/garfinkl/web/mog/} 
All issues before number {\bf 28} are available at
\htmladdnormallink {\protect {\tt {http://www.phys.lsu.edu/mog}}}
{http://www.phys.lsu.edu/mog}

Any ideas for topics
that should be covered by the newsletter, should be emailed to me, or 
Greg Comer, or
the relevant correspondent.  Any comments/questions/complaints
about the newsletter should be emailed to me.

A hardcopy of the newsletter is distributed free of charge to the
members of the APS Topical Group on Gravitation upon request (the
default distribution form is via the web) to the secretary of the
Topical Group.  It is considered a lack of etiquette to ask me to mail
you hard copies of the newsletter unless you have exhausted all your
resources to get your copy otherwise.

\hfill David Garfinkle 

\bigbreak

\vspace{-0.8cm}
\parskip=0pt
\section*{Correspondents of Matters of Gravity}
\begin{itemize}
\setlength{\itemsep}{-5pt}
\setlength{\parsep}{0pt}
\item Daniel Holz: Relativistic Astrophysics,
\item Bei-Lok Hu: Quantum Cosmology and Related Topics
\item Veronika Hubeny: String Theory
\item Pedro Marronetti: News from NSF
\item Luis Lehner: Numerical Relativity
\item Jim Isenberg: Mathematical Relativity
\item Katherine Freese: Cosmology
\item Lee Smolin: Quantum Gravity
\item Cliff Will: Confrontation of Theory with Experiment
\item Peter Bender: Space Experiments
\item Jens Gundlach: Laboratory Experiments
\item Warren Johnson: Resonant Mass Gravitational Wave Detectors
\item David Shoemaker: LIGO Project
\item Stan Whitcomb: Gravitational Wave detection
\item Peter Saulson and Jorge Pullin: former editors, correspondents at large.
\end{itemize}
\section*{Topical Group in Gravitation (GGR) Authorities}
Chair: Manuela Campanelli; Chair-Elect: 
Daniel Holz; Vice-Chair: Beverly Berger. 
Secretary-Treasurer: James Isenberg; Past Chair:  Patrick Brady;
Members-at-large:
Laura Cadonati, Luis Lehner,
Michael Landry, Nicolas Yunes,
Curt Cutler, Christian Ott,
Jennifer Driggers, Benjamin Farr.
\parskip=10pt

\vfill
\eject

\vfill\eject

\section*{\centerline
{we hear that \dots}}
\addtocontents{toc}{\protect\medskip}
\addtocontents{toc}{\bf GGR News:}
\addcontentsline{toc}{subsubsection}{
\it we hear that \dots , by David Garfinkle}
\parskip=3pt
\begin{center}
David Garfinkle, Oakland University
\htmladdnormallink{garfinkl-at-oakland.edu}
{mailto:garfinkl@oakland.edu}
\end{center}

Frans Pretorius has received a Simons Investigator Award.

Beverly Berger was elected Vice Chair of GGR; 
Benjamin Farr, Curt Cutler, and Christian Ott were elected Members at large of the Executive Committee of GGR.

Hearty Congratulations!

\section*{\centerline
{100 years ago}}
\addtocontents{toc}{\protect\medskip}
\addcontentsline{toc}{subsubsection}{
\it 100 years ago, by David Garfinkle}
\parskip=3pt
\begin{center}
David Garfinkle, Oakland University
\htmladdnormallink{garfinkl-at-oakland.edu}
{mailto:garfinkl@oakland.edu}
\end{center}

In 1912 Einstein continues to develop a gravitational theory where space is flat, but time is warped through
a spatially dependent $c$.  He proposes a Poisson type equation for $c$, but then modifies it by adding a 
nonlinear term to take into account ``the energy density of gravitation itself."  He also notes that the 
equations of motion for free fall particles in this theory can be derived from the same variational principle as
in special relativity.  (See Annalen der Physik {\bf 38} 355-369 and 443-458). 

\section*{\centerline
{New publisher and new book}}
\addtocontents{toc}{\protect\medskip}
\addcontentsline{toc}{subsubsection}{
\it New publisher and new book, by Vesselin Petkov}
\parskip=3pt
\begin{center}
Vesselin Petkov, Minkowski Institute
\htmladdnormallink{vpetkov-at-minkowskiinstitute.org}
{mailto:vpetkov@minkowskiinstitute.org}
\end{center}

A new academic publisher, the Minkowski Institute Press, has been launched.  Its first book is 
Hermann Minkowski, Space and Time: Minkowski's papers on relativity (Minkowski Institute Press, Montreal 2012), 123 pages.
Minkowski's three papers have never been published together either in German or English and Das Relativit\"{a}tsprinzip has not been translated into English so far. 

More information about the publisher is available at

http://minkowskiinstitute.org/mip/

while more information about the book the book can be found at

http://minkowskiinstitute.org/mip/books/minkowski.html

\vfill\eject

\section*{\centerline
{Dark Matter News: Tentative Evidence of a 130 GeV}\\ \centerline{Gamma-Ray Line from Dark Matter Annihilation}  \\ \centerline{at the Fermi Large Area Telescope}}
\addtocontents{toc}{\protect\medskip}
\addtocontents{toc}{\bf Research briefs:}
\addcontentsline{toc}{subsubsection}{
\it Dark Matter News, by Katherine Freese}
\parskip=3pt
\begin{center}
Katherine Freese, University of Michigan
\htmladdnormallink{ktfreese-at-umich.edu}
{mailto:ktfreese@umich.edu}
\end{center}

The Milky Way, along with other galaxies, is well known to be
encompassed by a massive dark matter (DM) halo of unknown composition.  
 A leading
candidate for this dark matter is a Weakly Interacting Massive Particle (WIMP).
The terminology refers to the fact that these particles undergo weak
interactions in addition to feeling the effects of gravity, but do not
participate in electromagnetic or strong interactions.  WIMPs are
electrically neutral.  A recent paper showed that, even with billions passing through our bodies
every second, on the average the number of interactions with the human body is at most one per minute
 \cite{Freese:2012rp}.
Their expected masses range from 1~GeV to 10~TeV.  Many WIMPs are their own antiparticles.
These particles, if present in thermal equilibrium in the early
universe, annihilate with one another so that a predictable number of
them remain today.  The relic density of these particles is
\begin{equation}
  \Omega_\chi h^2 \sim (3 \times 10^{-26} \mathrm{cm}^3/\mathrm{sec})
                        / \langle \sigma v \rangle_{\mathrm{ann}}
\end{equation}
where $\Omega_{\chi}$ is the fractional contribution of WIMPs
to the energy density of the Universe, and  $\langle \sigma v \rangle_{\mathrm{ann}}$ is the product of 
annihilation cross section times velocity.  A value of $\langle \sigma v \rangle_{\mathrm{ann}}$
of weak interaction strength automatically gives the right answer for the relic density, near
the value measured by WMAP \cite{Komatsu:2010fb}.
This coincidence is known as the ``WIMP miracle'' and is why
WIMPs are taken so seriously as dark matter candidates.  Possibly the
best WIMP candidate is motivated by Supersymmetry (SUSY): the lightest
neutralino in the Minimal Supersymmetric Standard Model (MSSM) and its
extensions~\cite{Jungman:1995df}.  However, other WIMP candidates arise
in a variety of theories beyond the Standard Model (see
Refs.~\cite{Bergstrom:2000pn,Bertone:2004pz} for a review).

\begin{figure}
	\includegraphics{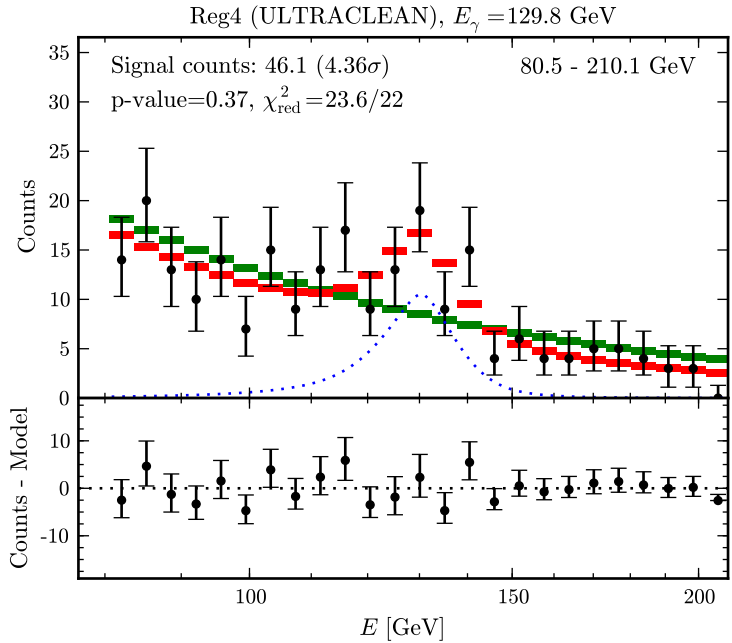}
  \caption{Figure taken from Weniger (arxiv::1204.2797) corresponding to a region near the Galactic Center.  
  Measured events with statistical
  errors are plotted in black.  The horizontal bars show the best-fit models with (red) and
  without DM (green); the blue dotted line indicates the corresponding line flux
  component alone.  The lower sub panel shows residuals after subtracting the model with
  line contribution. }
  \label{fig:line}
\end{figure}

A multitude of experimental efforts are currently 
underway to detect WIMPs, with some claiming hints of detection.  There
is a three-pronged approach: particle accelerator, indirect detection
(astrophysical), and direct detection experiments.

The latest dark matter news is in the area of indirect detection.
The same annihilation process that these particles undergo in the early Universe
may further take place in the current Universe in areas of high dark matter density.
Wherever there is a large abundance of such WIMPs, they annihilate among themselves
into a variety of other particles, which eventually fragment and decay into gamma-rays,
positrons, and neutrinos.  All of these annihilation products are being searched for.

The FERMI satellite has been searching for gamma-rays such as those that
might be produced by dark matter annihilation or decay.  
Particularly interesting places to look in the Galaxy are the regions of expected high
dark matter abundance, including the Galactic Center and satellite dwarf galaxies.
While the FERMI collaboration
itself has released only bounds on dark matter, others have examined the data
to look for signatures.  A particularly interesting result has recently been released by
Weniger \cite{Weniger:2012tx}, who finds tentative evidence for a 130 GeV gamma-ray
line in a region
close to the Galactic Center (GC).  Such a line
would be produced if two WIMPs annihilated directly to two photons, each of which
has the same energy as the mass of the incoming WIMP.  Weniger finds that the
significance of the result is 4.6$\sigma$, or when taking into account the look-elsewhere 
effect, 3.2$\sigma$.  Other authors have pointed out that
annihilation to two photons is likely to be accompanied by annihilation to a photon
and a Z; thus there may instead be two lines and these authors find that again such an interpretation is consistent
with the data \cite{Rajaraman:2012db,Su:2012ft, Su:2012zg}.  Based on constraints on a continuous spectrum of photons that should accompany the line, the authors of Ref.
\cite{Cohen:2012me} argue that neutralinos cannot be an explanation for the line.
See \cite{Tulin:2012uq} and \cite{Bai:2012qy}
for models that may accommodate thermal dark matter.
Many theoretical models have been proposed to explain the 130 GeV line; as many as
53 papers already cite the original Weniger result.

This result is as yet tentative.  Since it is based on only 50 photons, further data will
be required, both from FERMI and from other upcoming gamma-ray experiments such as 
HESS-II, CTA, and GAMMA-400 \cite{Bergstrom:2012vd}.  In addition, the result has not yet been vetted by the FERMI collaboration.  Puzzling is also the fact that another 130 GeV line appears in the direction of the bright limb at Earth's horizon,
dominantly produced by cosmic ray showers in the atmosphere;  this cannot be explained by a dark matter signal. 
While it is encouraging that such a line is not seen throughout the data, e.g. not in the Galactic Plane away from
the GC, still
these limb events are perplexing.
It will be very interesting to see whether this tentative
hint of a 130 GeV gamma-ray line towards the Galactic Center persists over the next few years.

In order for the physics community to be persuaded that the dark matter particle
has been discovered, it will have to appear in more than one experiment.  As yet there is no evidence for
a 130 GeV particle other than in the gamma-ray line. As mentioned above, other than indirect searches for dark matter
annihilation, the other two methods for dark matter searches are direct detection and particle accelerators.
 Direct detection experiments, which seek to measure the
energy deposited by the elastic scattering of a WIMP from the Galaxy off of
 a nucleus in the detector, have seen anomalous events
which may be due to WIMPs, but not at 130 GeV masses.  DAMA, CoGeNT, and CRESST all have events possibly compatible
with $\sim$ 10 GeV WIMPs, though these results are in tension with null results from CDMS and XENON. DAMA data may also be compatible with 80 GeV WIMPs, although this region is almost certainly ruled out
by null results from CDMS and XENON.  
The Large Hadron Collider at CERN could in principle detect a 130 GeV particle,
but has not seen any evidence.  Indeed some of 
the theoretical models for the 130 GeV line mentioned above
may not lead to signatures at the LHC at all.
For example, if the WIMP only couples to photons, then it would not be produced
in the collisions of two protons at the accelerator.    Nonetheless it is certainly possible that all three prongs of the experimental
searches for dark matter will provide future tests of the dark matter interpretation of the 130 GeV line.

\vfill\eject

\section*{\centerline
{Successful First Launch of ESA's VEGA Booster}\\ \centerline{Carries LARES Satellite into Perfect Orbit}}
\addtocontents{toc}{\protect\medskip}
\addcontentsline{toc}{subsubsection}{
\it LARES satellite, by Richard Matzner}
\parskip=3pt
\begin{center}
Richard Matzner, University of Texas
\htmladdnormallink{matzner2-at-physics.utexas.edu}
{mailto:matzner2@physics.utexas.edu}
\end{center}

Just before dawn in the morning of Monday, 13 February  2012 the European Space Agency (ESA) successfully carried out the first launch of its booster VEGA from the ESA launch site at Kourou in French Guiana. VEGA flight 01 (VV01) carried multiple scientific payloads; the principal one was the laser ranged LARES.  LARES (Laser Relativity Satellite) is a small (radius $182 mm$), massive ($386.8 kg$), passive laser-ranged satellite. The laser ranging is accomplished via $38.1 mm$ cube corner retroreflectors ($92$ total) set in latitude rows on the satellite. Except for the retroflectors and their mounting hardware, LARES is entirely made from a single piece of sintered tungsten, giving it a volume density of approximately $18 g/cm^3$, and a cross sectional area/mass ratio of $370g/cm^2$. It is by far the densest known orbiting body in the solar system. Laser observation of LARES on 17 February produced the values: semimajor axis $7828km$; orbital eccentricity $9 \times 10^{-4}$; inclination $69^\circ .4$; an ideal orbit to test a gravitational effect predicted by Einstein's General Relativity.

LARES is the (evolved) embodiment of an idea proposed by I. Ciufolini (University of Salento, now at University of Rome) in his PhD dissertation at The University of Texas at Austin in the mid- 1980s, to measure the Lense-Thirring effect (frame dragging effect) of the Earth, as a complement to the then-proposed gyroscope experiment Gravity Probe- B (GP-B). (John A. Wheeler and I were Ciufolini's co-advisors.) The rotation of the earth slowly ``drags space" with its rotation, and satellite orbits are dragged in the direction of the rotation compared to the Newtonian prediction, and more importantly compared to the reference frame defined by the distant stars. The expression for the Lense-Thirring frame- dragging rate $\Omega_{LT}$ is\\
\begin{equation} 
\Omega_{LT}= \frac{2 G J}{c^2a^3(1-e^2)^\frac{3}{2}},\\
\label{} 
\end{equation} 
\medskip

\noindent where $J$ is the angular momentum of the Earth, $a$ is the semimajor radius and $e$ is the eccentricity of the satellite orbit; $G$ is the gravittional constant and $c$ is the speed of light. 
For mid-range Earth orbits this dragging amounts to meters per year, while the tracking accuracy via laser ranging is in the millimeter range. Hence LARES will (contribute to) measuring the frame dragging effect of the Earth. (The GP-B experiment measured the spin axis direction of gyroscopes by comparing to a specified guide star.)

Above I used the construct ``contribute to" because no single satellite orbit can determine the frame dragging rate. The obstacle in the orbit tracking method is that Newtonian gravitational effects from the nonspherical Earth cause orbit-plane precessions that are up to $10^8$ times faster than the frame dragging precession. However the Newtonian effect is a function of the inclination of the orbit (which is measured from $0^\circ$ for a prograde orbit in the equatorial plane, to $180^\circ$ for a retrograde equatorial orbit) in contrast to the frame dragging rate, which is independent of the orbital inclination. The current method using laser-ranged satellites is to obtain the best extant description of the Earth's gravitational field, encapsulated in its multipole expansion (the best being derived from the GRACE observations (F\"orste et al. (2008a,b)), see also {\tt http://icgem.gfz-potsdam.de/ICGEM/ICGEM.html}); the even-order harmonics are responsible for the secular Newtonian precession, and to combine the observations of a number of laser-ranged satellites with different inclinations, the best being LAGEOS (launched in 1976, inclination $109^\circ .8$), LAGEOS-2 (1992, inclination $52^\circ .6$), and now LARES.  LAGEOS and LAGEOS-2 are in near-circular orbits of radius $12,270 km$ and $12,163 km$ respectively.

GRACE provides highly accurate determinations of fairly high order gravitational multipoles. It works by ranging between two identical drag-free satellites about $220$ kilometers apart in a polar orbit of $500$ kilometers altitude (orbit radius about $6900km$). The Newtonian secular effect of gravitational multipoles decreases as the order increases so the lowest few multipoles have the most significant effect on the precession of orbits. At LAGEOS orbital radius only $J2$ and $J4$ contribute significantly to the Newtonian precession. To determine the Lense-Thirring dragging, one can view the process as follows: GRACE determines the $J4$ and higher multipoles, and LAGEOS and LAGEOS-2 determine the Lense-Thirring dragging and the $J2$ multipole. (Actually all the variables are determined simultaneously, and the LAGEOS satellites contribute a $correction$ to the GRACE $J2$.) In this way one can determine a $10\%$ validation of the General Relativity Lense-Thirring dragging (Ciufolini and Pavlis, 2004; Ries et al., 2008; Ciufolini et al., 2009).  

The $10\%$ error is a systematic error arising mostly from uncertainty in the Earth's multipoles, evaluated in a root-mean-square analysis.  For comparison the GP-B experiment quotes a $19\%$ systematic error arising mostly from uncertain torques on the gyroscopes (Everitt et al. 2011), also evaluated in a root-mean-square analysis. With a bit of astrophysical skepticism, one can fairly say these errors ($10\%$, $19\%$) are comparable. 

With a long enough history (at least five years) of tracking LARES its orbit will be well enough known that the frame dragging determination can be redone including it. Then conceptually LARES, LAGEOS, and LAGEOS-2 can determine the frame dragging and the $J2$ and $J4$ harmonics while the GRACE-derived Earth gravity models provide the $J6$ and higher gravitational multipoles. Analysis predicts that the frame-dragging systematic error in that case will be about one order of magnitude better than in the case of the LAGEOS and GP-B results  (still mostly due to uncertainty in GRACE-derived model errors), a significant improvement on the current situation.

The LARES theory and data-analysis group includes I. Ciufolini, (University of Rome, University of Salento and INFN Sezione di Lecce, Italy), E. C. Pavlis, (University of Maryland at Baltimore County),  J. C. Ries  and Richard A. Matzner (University of Texas at Austin),  A. Paolozzi, (Sapienza University, Roma Italy), R. K\"onig (GFZ German Research Centre for Geosciences, Potsdam, Germany), Victor J. Slabinski (US Naval Observatory), and G. Sindoni (Sapienza University, Roma Italy).    \\

\bigskip

\begin{center}
{\bf References}
\end{center}

\noindent Ciufolini, I., Pavlis, E.C., 2004. $Nature$ {\bf 431}, 958.\\

\noindent Ciufolini, I., Paolozzi, A., Pavlis, E.C., Ries, J.C., K\"onig, R., Matzner, R.A., Sindoni, G., Neumayer, H., 2009. {\it Space Sci. Rev.} {\bf 148}, 71.\\

\noindent Everitt, C. W. F.  et al. (2011). {\it Physical Review Letters} {\bf 106}, 221101.\\ 

\noindent F\"orste, C., Flechtner, F., Schmidt, R., Stubenvoll, R., Rothacher, M., Kusche, J., Neumayer, K.-H., Biancale, R., Lemoine, J.-M., Barthelmes, F., Bruinsma, J., K\"onig, R., Meyer, U., 2008a. EIGEN-GL05C Ð A new global combined high resolution GRACE-based gravity field model of the GFZ-GRGS cooperation. General Assembly European Geosciences Union (Vienna, Austria 2008), {\it Geophysical Research Abstracts}, {\bf 10}, No. EGU2008-A-06944.\\

\noindent F\"orste, C., Schmidt, R., Stubenvoll, R., Flechtner, F., Meyer, U., K\"onig, R., Neumayer, H., Biancale, R., Lemoine, J.-M., Bruinsma, S., Loyer, S., Barthelmes, F., Esselborn, S., 2008b. {\it J. Geodesy} {\bf 82} (6), 331. \\

\noindent Ries, J.C., Eanes, R.J., Watkins, M.M., 2008. ``Confirming the frame-dragging effect with satellite laser ranging". In: {\it 16th International Workshop on Laser Ranging}, 13Ð17, Poznan, Poland. {\tt http://cddis.gsfc.nasa.gov/lw16/docs/presentations/sci\_3\_Ries.pdf} and {\tt http://cddis.gsfc.nasa.gov/lw16/docs/papers/sci\_3\_Ries\_p.pdf}.\\

\vfill\eject
\section*{\centerline
{Workshop on Gravitational Wave Bursts}}
\addtocontents{toc}{\protect\medskip}
\addtocontents{toc}{\bf Conference reports:}
\addcontentsline{toc}{subsubsection}{
\it Workshop on Gravitational Wave Bursts, 
by Pablo Laguna}
\parskip=3pt
\begin{center}
Pablo Laguna, Georgia Tech 
\htmladdnormallink{plaguna-at-gatech.edu}
{mailto:plaguna@gatech.edu}
\end{center}

An unprecedented
view of the explosive and transient gravitational-wave sky will be available by the end of this decade, thanks to interferometric detectors. The time is ripe to challenge our
  theoretical understanding of short duration gravitational-wave
  signatures from cataclysmic events, their connection to more
  traditional electromagnetic and particle astrophysics, and the data
  analysis techniques that will make the observations a reality.  The workshop series
 on \emph{Gravitational Wave Bursts:
  Astrophysics, Data Analysis and Numerical Relativity} have been conceived with such objectives in mind, bringing together, in a
remote and inspiring location, scientists in astrophysics, data
analysis and numerical relativity to discuss, analyze and explore
innovative views on the \emph{Transient Gravitational Wave
  Universe}. These workshops emphasize discussion over
presentations, with a format designed to encourage conversations and
critical evaluation of efforts and methodologies. Because of its modest number of participants,
the workshops provide a natural vehicle that promotes synergistic
collaborations.

The first \emph{GWburst} workshop took place in
Chichen-Itza, M\'exico during December 9-11, 2009 ({\tt http://gwbursts.org/}).
The second workshop took place during May 28-30, 2012 in the
small fishing port of Tobermory, on the Isle of Mull, off the west
coast of Scotland ({\tt http://www.tobermory.co.uk/}).
Discussion topics were:
\begin{itemize} \setlength{\itemsep}{-2pt}
\item Astrophysics behind GWburst sources (e.g. stellar core collapse,
  gamma-ray bursts, cosmic strings, compact object mergers, isolated
  neutron stars) and their connection with electromagnetic and
  neutrino observations.
\item Challenges to numerically model transient sources and the
  required accuracy of simulations.
\item Data analysis methodologies to detect and characterize GWbursts.
\item Gravitational wave antennas and their capabilities.
\item Detection of unknown GWburst sources.
\end{itemize} 

This year's workshop was organized around the following sessions:
\begin{itemize} \setlength{\itemsep}{-2pt}
\item Core-Collapse SNe and Long GRBs,
\item NS/NS, NS/BH Merger and Short GRBs,
\item Isolated Neutron Stars,
\item Binary Black Holes,
\item Data Analysis, and
\item Instrumentation. 
\end{itemize}

Each session started with a {\it Mano-a-Mano} discussion in which two invited speakers provided not only their  broad view of the field (e.g. Core-Collapse SNe and Long GRBs, Data Analysis, Instrumentation, etc) but they also gave what in their opinion were future directions and what could be done better.  They were encouraged to in particular focus on controversial subjects,  open questions and key challenges.  An example of a topic that triggered a passionate conversation/discussion was regarding open data. The {\it Mano-a-Mano} discussion was followed by
presentations on specific topics  (e.g. Are GRBs powered by magnetars, Numerical simulations of eccentric NS binaries, etc) and open discussions, with the sessions ending with a summary by the chairs. 
The workshop website, which includes links to the talks and summary discussions, can be found at {\tt
  http://www.physics.gla.ac.uk/igr/GWbursts2012}.
  
  As with the first workshop, there was a general consensus that the ``formula'' for these meetings is highly successful and thus that the {\it GWburst} workshop series must be continued.
 With Advanced LIGO and Advanced Virgo around the corner, understanding the {\it Transient Gravitational-wave Sky} requires a conversation among astrophysicists, numerical relativists and data analysts that workshops such as this enable. 

\vfill\eject
\section*{\centerline
{JoshFest}}
\addtocontents{toc}{\protect\medskip}
\addcontentsline{toc}{subsubsection}{
\it JoshFest, 
by Ed Glass}
\parskip=3pt
\begin{center}
Ed Glass, University of Michigan 
\htmladdnormallink{englass-at-umich.edu}
{mailto:englass@umich.edu}
\end{center}

2012 marks the fiftieth anniversary of the publication of the famous
"Goldberg-Sachs theorem" and the sixtieth anniversary of Josh Goldberg's PhD award.

The day of the Joshfest, Friday, April 20th, was unseasonably warm for
Syracuse. The physics department hosted the celebration with invited talks
starting in the early afternoon. The chairman, Peter Saulson gave a short
intro and then your reporter spoke about how the GRG volume came to be (volume
43 number 12). The volume contains 25 relativity articles and an introduction
by David Robinson and Ed Glass. Josh was presented with an Einstein flash
drive and the bespoke volume.

The presentation was followed by talks given by John Stachel, Peter Saulson,
Rafael Sorkin, and Mark Trodden. The talks had many photos of relativists from
the "heroic" era, with some of those relativists in the audience. After the
talks everyone walked across campus to a private home long since renovated as
a faculty club where there were drinks and dinner. After dinner Peter Saulson
read a few emails from people who couldn't attend but sent greetings (the
modern version of telegrams). Ted Newman, the famous comedian, told an
assortment of lies about Josh's undergraduate career.

Josh spoke briefly about how much he enjoyed the day. His talk was followed by
spontaneous remarks from people in the dinner audience. They all said,
variously, that the congeniality, warmth, and hospitality of the Syracuse
physics department was due to the personal efforts of Josh and his wife Gloria.
\vfill\eject

\section*{\centerline
{Connecting the Electromagnetic and Gravitational Wave Skies}\\ \centerline{ in the Era of Advanced LIGO}}
\addtocontents{toc}{\protect\medskip}
\addcontentsline{toc}{subsubsection}{
\it Electromagnetic and Gravitational Wave Astronomy, 
by Sean McWilliams}
\parskip=3pt
\begin{center}
Sean McWilliams, Princeton University 
\htmladdnormallink{stmcwill-at-princeton.edu}
{mailto:stmcwill@princeton.edu}
\end{center}

The Princeton Center for Theoretical Science (PCTS) hosted a 5-day workshop entitled ``Connecting the Electromagnetic and Gravitational Wave Skies in the Era of Advanced LIGO" from April 30 to May 4, 2012.  Organized by Adam Burrows, Sean McWilliams, Brian Metzger, Frans Pretorius, David Spergel, Anatoly Spitkovsky and Paul Steinhard, this workshop brought together members of the gravitational wave (GW) and electromagnetic (EM) transient communities to discuss theoretical and observational questions of common interest.  Advanced LIGO is expected to begin taking science data in early 2015, and to reach its design sensitivity a few years thereafter.  Likewise, transient electromagnetic events are a major focus of ongoing astronomical study, with Swift and Fermi responding rapidly to gamma-ray burst events and various other wide-field telescopes (e.g. LOFAR in radio, LSST in optical) in existence or planned for the near future.  Many theoretical models predict coincident GW and EM emission from compact binary coalescences and supernovae, so coordination between the two communities will be critical to maximize the scientific payoff of observing these sources.

Day 1 of the workshop focused on the status of Advanced LIGO commissioning, the plans for science runs beginning in 2015, the schedule for incremental development as Advanced LIGO approaches its design sensitivity, and plans for missions in the more distant future.  Day 2 focused on specific predictions for source populations and event rates for Advanced LIGO, the status of source modeling, and the achievable accuracy for measuring source parameters through GWs.  The topic of Day 3 was the status of numerical simulations of binary systems, including incorporating matter and electromagnetic fields, generating and transporting neutrino and photon radiation, and taking relevant microphysics into account.   Day 4 transitioned to the GRB-merger connection, including coordinated GW-GRB observational strategies and a discussion of the implications of known GRBs for future GW observations.  Finally, Day 5 concluded with a discussion of other potential EM signatures of merger events across the EM frequency band, their relative likelihoods, strategies used during initial LIGO for EM followup, and potential strategies for EM followup in the Advanced LIGO era.

Further details, including slides and recordings of each presentation and topic summaries for each discussion panel, can be found at http://pcts.wikispaces.com.

\vfill\eject
\section*{\centerline
{KITP Program report: Bits, Branes, and Black Holes}}
\addtocontents{toc}{\protect\medskip}
\addcontentsline{toc}{subsubsection}{
\it Bits, Branes, and Black Holes, 
by Ted Jacobson and Don Marolf}
\parskip=3pt
\begin{center}
Ted Jacobson, University of Maryland 
\htmladdnormallink{jacobson-at-umd.edu}
{mailto:jacobson@umd.edu}
\end{center}
\begin{center}
Don Marolf, University of California, Santa Barbara
\htmladdnormallink{marolf-at-physics.ucsb.edu}
{mailto:marolf@physics.ucsb.edu}
\end{center}

For ten weeks this spring (March 19 - May 25, 2012),
the Kavli Institute of Theoretical Physics (University of California, Santa Barbara) held a
focussed program addressing i) the classic question of whether and how information escapes from black holes and ii) the nature and scope of ``holography'' as typified by the AdS/CFT correspondence of string theory.  Although there was an emphasis on string theoretic techniques, participants in the program represented a broad range of approaches and expertise.

There were roughly 45 seminars and scheduled discussions (sometimes called working groups)
as well as 30 half-hour talks and 4 panel discussions in the concluding conference.
Most of the discussions lasted close to 2 hours, and usually involved
one or more presentations, followed or interrupted by open discussion.
Here we will attempt to summarize the themes that ran through much of the program,
and to give a brief guide to the various talks and discussions.

Most of the official events were recorded (both audio and video).
Links to these recordings, as well as some slides, notes and reference materials
can be found at the Program Wikispace, http://bitbranes12.wikispaces.com/
and the Conference Archive, http://online.kitp.ucsb.edu/online/bitbranes-c12/
\\

\noindent  {\bf 1) The black hole information question:}
In 1976 Hawking concluded that
since information seems to be lost forever inside a black hole,
there can be no unitary S-matrix for the process of black-hole formation and evaporation.
Debate has continued ever since.
While viewpoints have tended to become rather entrenched over time,
discussion at the program was wide open,
and a number of participants professed to be
less sure of their viewpoint than previously.
The current view expressed by
nearly all participants is that
arguments from gauge-gravity duality and canonical quantum gravity
at least strongly suggest that in fact a unitary black hole S-matrix must exist.
(Arguments that information loss would necessarily entail
copious black hole pair creation and violations of energy and momentum
conservation were also reprised, but are not as widely accepted.)
This leads to what is variously
called the black hole information ``problem", ``puzzle", or
``paradox": how can information \emph{not} be lost, given that apparently well-justified
semiclassical reasoning says that it is lost? Discussion of black hole information
at the program tended to focus on this puzzle.

The program began with a summary of the question and a review of some
points and counterpoints that have been made in the past,
presented by Ted Jacobson and Joe Polchinski (who also gave a conference
talk on his current viewpoint).
In a program seminar and a conference talk, Samir Mathur
advocated a ``fuzzball" picture of black hole microstates,
deduced in a string theory setting, arguing that it implies there
is no black hole interior into which information can be lost.
In a program seminar and a conference talk
Steve Giddings advocated that a radical nonlocal modification of physics
is required to account for black hole unitarity,
and he presented simple models designed to explore this possibility.
In one discussion Don Marolf presented an argument, depending only on general
covariance, in favor of the existence of a unitary black hole S-matrix.
The crux of the argument is that, since the total Hamiltonian
lives at spatial infinity, the observables at infinity must evolve unitarily.
In another discussion Bill Unruh presented a model showing how decoherence
requires no energy transfer, and he used this to argue that
exterior information loss need not entail violation of energy conservation.
Other discussions focused on what was learned from 1+1 dimensional models
(led by Steve Giddings)
(this was inconclusive, since the models are dynamically incomplete),
what is known about ``fast scrambling" in quantum systems that could be
models for black hole dynamics (with program presentations by Jose Barbon and Nima Lashkari and conference talks by Patrick Hayden and Douglas Stanford),
and a panel discussion ``Black hole information"
at the conference (chaired by Gary Gibbons, with panelists
Steve Giddings, Samir Mathur, Joe Polchinski and Rafael Sorkin).\\

\noindent {\bf 2) Horizon entropy and microstates:}
The black hole information question is closely related to that
of the statistical meaning of the entropy of black holes and other horizons,
which in turn hinges on the nature of the microstates that the entropy might
be counting.  This remains an outstanding question, forty years after
Bekenstein's bold proposal that black holes have entropy proportional to
the horizon area in Planck units. A large number of talks and discussions
were devoted to this topic.

One lively discussion session focused on Mathur's fuzzball proposal,
involving classical solutions with compact higher dimensions
in which spacetime is closed off by topologically nontrivial structure
near where a horizon would be. Part of the focus was on
how this picture could be compatible with semiclassical physics
where that applies. Mathur proposed that this picture applies even to
Rindler acceleration horizons in flat spacetime, which could be
thought of as quantum superpositions of correlated fuzzball states.
This invoked the notion, advocated by Mark van Raamsdonk in
a seminar and a conference talk, that AdS/CFT duality implies that
spatial continuity in bulk physics amounts to superpositions
of singular, correlated but disconnected geometries.  Nick Warner and Jan de Boer also gave their own takes on the fuzzball program in their conference talks.  While Warner emphasized the many explicit classical supergravity fuzzball solutions found to date,  de Boer reinforced the argument that classical supergravity solutions alone cannot account for black hole entropy.  However, he suggested that so-called non-geometric solutions of classical string theory could play an important role.

Another discussion concerned what part of the phase space of
gravity (with a negative cosmological constant) is dual to states
in the CFT, and included presentations by Ted Jacobson, Steve Hsu,
Steve Giddings and Alex Maloney as well as much open discussion.
In particular Hsu explained how in classical GR one can construct apparently
compact objects with fixed ADM mass but arbitrarily large entropy.
These objects collapse into black holes but have more entropy
than the area of the resulting black hole, so apparently could not
be dual to states in the CFT.
It seems that either some of the phase space is
excluded, or it doesn't survive quantization. Maloney discussed a
calculation in 2+1 dimensional quantum gravity illustrating how
classical configurations can fail to survive quantization.

Three discussions (not recorded)
concerned the nature, size, and
role of quantum horizon fluctuations, the final one
focusing on the question of whether the notion of horizon
remains meaningful beyond the semiclassical setting. Several
arguments were given on both sides, with most participants
taking the view that the concept of horizon does not
survive beyond perturbative quantum gravity.
In a seminar and a conference talk, Fay Dowker
discussed properties and assessed the promise
of a proposed definition of horizon entropy
in the setting of discrete causal sets. The definition was
the difference
between the sum of the (non-local) causal set actions of
the two sides of the horizon and the action of their union.

The properties of entanglement entropy of quantum fields
and its relation to black hole entropy were the focus of
quite a few talks and discussions. In one discussion
William Donnelly reviewed the longstanding question
of the effects of nonminimal coupling to curvature in the
one-loop contribution of matter fields to black hole entropy.
In particular he argued against the validity of
previous calculations showing
that gauge fields, because of nonminimal coupling in the
gauge-fixed action, contribute negatively to black hole entropy.
In another discussion and in a conference talk,
motivated also by the causal set paradigm, Rafael Sorkin
presented a formulation of horizon entropy for a spacetime
region, and a formula for the entanglement entropy of
a free scalar field in terms of a nonlocal expression involving
nothing but the 2-point function in that region.
Sergey Solodukhin presented a formula for the average
Renyi entropy across a surface
of area A in a spacetime that is Minkowski times a compact
2d space with arbitrary geometry.
Holographic entanglement entropy, an approach to computing and using
this quantity in the AdS/CFT context, is discussed in the section ``Decoding
Holography" below.

Recent progress in the computation of black hole entropy
in the setting of string theory and beyond was presented by Atish Dabholkar
in a seminar and a conference talk, and by Ashoke Sen in a conference talk.
Dabholkar discussed systems in which exact evaluation of the
field theoretic functional integral for the
partition function of BPS (extremal, supersymmetric) states can be carried
out and compared with microscopic state counting. He also described
how the ``mock modular forms" mentioned in a l920 letter from
Ramanujan to Hardy arise in this context,
which provides an infinite
number of previously unknown such forms.
Sen's talk concerned mainly the contribution of massless fields to
the entropy involving the logarithm of horizon area.
He presented both extremal supersymmetric cases where the result can
be directly compared with microscopic state counting, and more general
results applicable even to a neutral, Schwarzschild black hole for example.

Other approaches to finding a microscopic understanding of black hole entropy were reviewed in program talks by Monica Guica and Mirjam Cvetic, and in the conference talk by Finn Larsen.  Cvetic and Larsen focussed on spacetimes with a so-called `hidden conformal symmetry' associated not with spacetime Killing fields but instead with properties of linearized fields on these spacetime backgrounds.  The approach relies on the deep connection between hypergeometric functions and the conformal group to look for signs of a dual conformal field theory.  In contrast, Guica's talk reviewed recent ideas for identifying a field theory dual to black holes that resemble nearly extreme Kerr.  This approach is known as the Kerr/CFT correspondence, and is notable for its lack of supersymmetry and its focus on describing more realistic black holes.

Finally, a discussion (unrecorded) involving
presentations by Jennie Traschen, Gary Gibbons and Alex Maloney
was organized around the topic of ``Horizon thermodynamics with a varying cosmological constant,
the inverse isoperimetric inequality, and the grand canonical ensemble
of quantum gravity".\\

\noindent {\bf 3) Higher Spin Holography:}  Most readers will have at least passing familiarity with Maldacena's AdS/CFT correspondence, the idea that string theory in asymptotically anti-de Sitter (AdS) ``bulk'' spacetimes is somehow equivalent to a (non-gravitational) conformal field theory in a smaller number of dimensions.  One obstacle to better understanding this correspondence is the fact that familiar gravitational physics arises in the bulk only in a strongly coupled limit of the dual CFT.  It is thus difficult to study this regime using conventional field theory techniques.

Some years ago, Klebanov and Polyakov  suggested that an analogous correspondence should hold for a class of ``free'' CFTs known as vector $O(N)$ models.  The simplest such model is just a set of $N$ free scalar fields (conformally coupled to the background metric), subject to the constraint that one restricts the operator algebra to those operators invariant under global $O(N)$ rotations.  E.g., one might consider $O = \sum_i \phi_i(x) \phi_i(y)$ even at separated points $x,y$, but not $\phi_i(x)$ itself.

One might expect that a simple CFT must be dual to a rather complicated bulk gravity theory.  In particular, Klebanov and Polyakov conjectured that this free theory (say, for $d=2+1$) is dual to a theory known as Vasiliev gravity in AdS${}_4$.   Some intuition behind this idea stems from the fact that any free theory (and even with the above constraint imposed) has an infinite number of conserved currents.  The stress tensor is a particular (spin 2) example, and the other currents form an infinite tower associated with higher and higher (even) spins.  The operator $\sum (\phi_i)^2$ can also be thought of as a spin zero current. In the usual gauge/gravity duality the CFT stress tensor is dual to the (spin 2) bulk graviton.  One might thus expect the free $O(N)$ vector model to be dual to some bulk theory containing an infinite tower of higher spin gauge fields (of all even spins $\ge 2$) as well as a spin zero field.  This is precisely the defining feature of  Vasiliev gravity, also called higher spin gravity.   The associated higher spin gauge symmetries act non-trivially on the metric, so that geometry (and even metric causal structure) are not gauge-invariant.  The theory is non-local in the sense that its equations of motion include an infinite number of derivatives (in both time and space), though when linearized about pure AdS space they become the usual two-derivative Fronsdal equations for the higher spin fields (and the usual results for the graviton and a conformally-coupled scalar).  The general structure of Vasiliev gravty was nicely reviewed by Wei Song in her seminar for the program.  In addition to the duality with the above-mentioned free theory, it turns out that versions of Vasiliev gravity can be dual to certain interacting theories as well.

Understanding and developing this new duality is a very active area of research that was the focus of program talks by Steve Shenker, Tom Hartman, and Juan Maldacena as well as conference talks by Xi Yin, Alejandra Castro, and Per Kraus.  The talks by Shenker, Hartman, Castro, and Kraus focused on matching CFT partition functions with bulk thermodynamics and in particular that of bulk black holes.  This is tricky since causal structure is not even gauge invariant in these theories!  In contrast, Maldacena and Yin looked at what this new example might teach us about gauge/gravity duality more generally. Might there be a sense in which every quantum field theory is dual to a suitably generalized notion of an AdS gravity theory?  Is there a sense in which all such bulk theories (even Vasiliev's) are string theories?  Readers looking for insight into these questions would do well to listen to the recorded talks.\\

\noindent {\bf 4) Dual formulations of de Sitter space \& Cosmology:}  It it natural to ask if some analogue of AdS/CFT can hold with a positive cosmological constant, or even in more general expanding cosmologies.  Several different approaches to this question have been developed and have made significant progress in the past few years.  The approaches represented in Bits, Branes, and Black holes are known as dS/CFT, the dS/dS correspondence, and acceleration from negative $\Lambda$.  The program also featured a seminar and discussion by Lenny Susskind concerning related work in which the structure of eternal inflation can give rise to a field theory (conformal or otherwise) at future infinity.

Let us begin with the dS/CFT correspondence, which may be thought of as AdS/CFT turned on its side (so that the timelike AdS boundary becomes the spacelike de Sitter boundary).   The idea is that the {\it arguments} of the Hartle-Hawking wavefunction of the universe (i.e., the 3-geometry and other fields) may be thought of as sources that one might couple to a Euclidean CFT.  The partition function $Z$ (as a function of these sources) is conjectured to give precisely the Hartle-Hawking wavefunction.

In principle, one might view this as a strict analytic continuation in $\Lambda$ of the AdS/CFT correspondence.  However, such an analytic continuation of string theory in AdS does not give string theory in dS, and instead gives a highly unstable theory.  Thus the standard AdS/CFT duality cannot simply be analytically continued to de Sitter space.  On the other hand, Anninos, Hartman, and Strominger recently showed that the above-mentioned dualities involving Vassiliev gravity are better behaved in this regard\footnote{This has to do with the fact that they contain no fields of odd integer spin.},  and that analytic continuation of the AdS theory to positive $\Lambda$ does indeed yield the de Sitter theory.  On the CFT side, analytic continuation can also be performed at each order in perturbation theory.  The results match the perturbation theory of a known CFT, which the authors then conjecture to supply the non-perturbative dual to de Sitter Vasiliev gravity.   This work was nicely reviewed in a program seminar by Andy Strominger and a conference talk by Dionysios Anninos.  However, as described in the conference talk by Dan Harlow, recent work suggests that the resulting bulk theory may still be unstable at the perturbative level.

In contrast, Eva Silverstein's program and conference talks
described the so-called dS/dS correspondence, which  builds on ideas that connect AdS/CFT with Randall-Sundrum (RS) braneworlds. In the usual version of AdS/CFT, the non-dynamical AdS boundary plays a central role.  In particular
excitations localized near the AdS boundary
have very high energy as defined by a fixed Killing field.  But the RS braneworld construction
effectively pushes the AdS boundary inward to a finite location and replaces it by a dynamical `brane'.  This both removes the above high-energy excitations and makes the boundary metric dynamical.  The result may be thought of as being dual to a CFT that has been cut-off at some (high) energy scale, and also coupled to a form of dynamical gravity associated with the dynamical boundary metric.
Dong, Horn, Silverstein, and Torroba have constructed analogous dualities in the de Sitter context by connecting the above picture
with constructions of meta-stable de Sitter vacuua in string theory. Though the CFT lives on a dynamical spacetime, this spacetime turns out to be close to $dS_{D-1}$ when the bulk is close to $dS_D$; thus the name ``dS/dS correspondence." These dS/dS scenarios have no explicit branes, but the RS brane's role as a cut-off is replaced by the spatial compactness of de Sitter space.  The rough counting of degrees of freedom in the dual field theory matches the de Sitter entropy, and generalizations exist for other expanding cosmologies.  Although different in many details, the approach described in Herman Verlinde's conference talk explored related ideas involving introducing covariant cut-offs on field theories in an attempt to build duals of Euclidean gravity on $S^4$, which is of course the Wick rotation of de Sitter.

Finally, Thomas Hertog's program seminar described work with Jim Hartle
and Stephen Hawking on how accelerated cosmologies can emerge from
quantum gravity
with negative $\Lambda$. They consider the semi-classical approximation to the Wheeler-DeWitt equation in theories of gravity coupled to scalar fields, and study these solutions at complex arguments  (i.e., on the space of complex 3-geometries and scalar fields).  This amounts to studying complex stationary points of the action. If the fundamental definition of the theory involves a $\Lambda$ of one sign and metrics of signature ($-$$+$$+$$+$), then stationary points of the opposite ($+$$-$$-$$-$) signature evolve as if governed by a $\Lambda$ of the opposite sign -- thus accelerating cosmologies from a theory whose fundamental definition involves a negative $\Lambda$.
Hartle and Hertog propose to use this result to make contact with AdS/CFT.  In particular, they hope to reverse this connection to use AdS/CFT to define the theory in the de Sitter context.\\

\noindent{\bf 5) Decoding Holography:}  Another
major theme of the program concerned the effort to make more precise the AdS/CFT dictionary relating observables deep in the bulk (perhaps even behind horizons) to observables in the CFT.  We now attempt to summarize some of the many approaches discussed at the program.

A particularly clear feature of the the AdS/CFT dictionary is the manner in which local CFT operators correspond to (rescaled) limits of bulk operators at AdS infinity.  It is therefore natural to extend this dictionary deeper into the bulk by attempting to solve the bulk equations of motion to express general bulk operators (perhaps perturbatively) in terms of their boundary values.  This is not the usual problem of Cauchy evolution, so a solution need not exist for generic boundary data.  But here we assume that we have `good' boundary data (supplied by the dual CFT) and ask whether the corresponding bulk solution is unique.  As described in program discussions by Gilad Lifschytz and Dan Kabat, at least in interesting circumstances, it turns out that it is and that this technique can indeed be used to reconstruct bulk operators.  For toy models involving only bulk scalar fields, it is known that the resulting bulk operators commute at spacelike separations.    This version of the bulk boundary map represents a given bulk observable as a complicated expression on the boundary, which is non-local in both time and space.  However, as described by Don Marolf in a program discussion,  one can in principle then go further by using the bulk Hamiltonian (represented as a boundary term) to localize this expression on a single Cauchy surface of the boundary spacetime.  Thus there is a sense in which any (perturbative) bulk observable can be written in terms of boundary data at a single boundary time and then mapped to a CFT observable at the same boundary time.

Furthermore, Joan Simon's program discussion described how, at least for solutions with enough supersymmetry, one can find rather more explicit ways to map bulk data to that of the CFT.  This discussion also addressed the CFT interpretation of certain timelike bulk singularities.

Rather than concentrate on smooth bulk fields, one might also try to use D-branes in the bulk.  This is particularly natural as D-branes are the basic building blocks of the AdS/CFT duality and one might expect them to be described simply in the CFT.  As described in program discussions by Gary Horowitz and Albion Lawrence, this is true in at least some sense:  the CFT has a so-called moduli space that describes the analogue of geodesic motion for the D-branes in the bulk.  One might thus hope that one can make use of this moduli space to describe bulk regions behind horizons. Discussion centered on the extent to which the moduli space coordinates are local in terms of the CFT fields.   In the same discussion session as Lawrence's presentation, Erik Verlinde also presented ideas that the CFT somehow contains `extra' degrees of freedom to describe black holes.

Another approach is to focus on the matrix-valued degrees of freedom associated with the $SU(N)$ symmetry in the field theory, following up an old idea that excitations localized to a few neighboring elements in this matrix are dual to excitations localized in the AdS bulk.
David Berenstein's conference talk reviewed one version of this approach which uses numerical simulations to study the CFT under the assumption that the CFT dynamics can be treated classically. One success of this program is that he finds at least qualitative agreement between the behavior of certain CFT sub-systems and that of localized `small' black holes in the bulk with size $R$ smaller than the AdS scale $\ell.$  Understanding these black holes is particularly important as they are the only ones relevant to the asymptotically flat $\Lambda \rightarrow 0$ limit of AdS.

The program seminar by Matt Headrick and the conference talk by Tadashi Takayanagi reviewed a final approach based on the so-called holographic entanglement entropy conjecture of Ryu and Takayanagi (and its time-dependent extension).  This (extended) conjecture proposes that
the (von Neumann) entropy in a CFT state restricted to a region $R$ in a
Cauchy surface is equal to the area (divided by $4G$) of
a certain co-dimension 2 surface in the corresponding bulk geometry, namely,
the extremal surface with minimal area that meets the AdS boundary at $\partial R$ and is homologous to $R$.
Tadashi Takayanagi discussed how this correspondence might be used to reconstruct the full bulk metric.  In addition, the program talks by both Mukund Rangamani and Mark van Raamsdonk discussed how the conjecture might be used to identify regions in the bulk dual to regions of the CFT.

Finally, two panel discussions at the conference addressed holography in general,
``Bulk Physics with CFT Duals" (Nick Warner (Chair), Frederik Denef, Finn Larsen, Ashoke Sen, Eva Silverstein),
and ``CFTs with Holographic Duals"
(Albion Lawrence (Chair), David Berenstein, Per Kraus, Shiraz Minwalla, Herman Verlinde).\\

\noindent {\bf 6) Other topics:} Many other interesting topics
which do not necessarily fit under the headings above
 were also discussed at the program.  While it is not possible to describe all of these in detail, we attempt to list them quickly so that interested readers can follow up by watching the corresponding talks on the web.\\

{\bf Black Holes, Hydrodynamics, and Blackfolds:}  A particularly useful outgrowth of AdS/CFT has been the development of the so-called fluid/gravity correspondence, which relates long-wavelength disturbances of planar black holes to solutions of the relativistic Navier-Stokes equations and their higher-order generalizations.  Mukund Rangamani's program seminar on this subject provides a lovely introduction to the subject and review of results to date.  Veronika Hubeny also gave a physics department colloquium on the subject, though unfortunately no recording is available.
Geoffrey Comp\`ere's conference talk addressed how this correspondence
can be set up near a Rindler horizon, imposing a Dirichlet
boundary condition for the metric on a surface at fixed distance from the
horizon.  Shiraz Minwalla's conference talk used this correspondence as a starting point to discuss his ideas using effective field theory methods
 to constrain hydrodynamics, and to relate the second law of thermodynamics to the existence of a partition function.  As noted in Rangamani's talk, the fluid/gravity correspondence is closely related to the so-called blackfold approach to black brane dynamics pioneered by Roberto Emparan and collaborators and reviewed in the program seminar by Niels Obsers and Jay Armas.\\

{\bf What other field theories have bulk AdS-like duals?:}  The program seminar by Kyriakos Papadodimas and the conference talk by Daniel Grumiller both addressed possible further generalizations of AdS/CFT.  Papadodimas reviewed ideas suggesting any large $N$ CFT with a small number of low-dimension operators should have an AdS-like gravity dual (with ``normal" as opposed to higher spin gravity).  Grumiller discussed possible extensions of AdS/CFT to so-called ``logarithmic'' CFTs (LCFTS).  LCFTS are non-unitary CFTs that often arise as limits of more familiar cases, in much the same way that differential equations develop logarithmic solutions at special values of their coefficients.\\

{\bf Explicit AdS/CFT calculations:}  In his program seminar, Balt van Rees reviewed how the introduction of Mellin transforms has enabled recent progress in understanding the 1/N expansion in the CFT, its connection to bulk perturbation theory, and to the extraction of an S-matrix in the $\Lambda \rightarrow 0$ limit of the bulk string theory.\\

{\bf The non-linear instability of AdS space:}  The conference talk by Jorge Santos reviewed recent results indicating that, at the level of classical GR,  at least certain open sets of small initial data near empty AdS space develop localized strong field regions which may evolve to black
holes. The work of Santos and collaborators builds on previous spherically symmetric results by Bizo\'n and Rostworowski.
This is in striking contrast with the non-linear stability of Minkowski space as shown by Christodoulou and Klainerman. The difference may be thought of as due to the fact that discrete spectrum of linearized modes in AdS contains large numbers of potential resonances that amplify the effect of non-linearities.\\

{\bf Spacetime Thermodynamics and Emergent Spacetime:}  It was argued some years ago by Ted Jacobson that the Einstein equation may be thought of as an equation of state.  Jacobson's conference talk reviewed this result and described attempts
and obstructions to extending the idea to more complicated gravitational theories that include higher derivative corrections.  Erik Verlinde's talks in both the program and the conference reviewed other ideas for thinking of gravitational dynamics as emergent thermodynamic behavior and the possible implications for understanding dark matter and cosmology.  On a related note,  Tom Bank's program talk (not recorded) discussed his Holographic Space-time proposal which also suggests that gravitational dynamics might emerge from a more fundamental structure -- this time one associated with a fixed non-dynamical spacetime causal structure.
Finally, at the conference there was a panel discussion on ``Emergence of Spacetime"
(Gary Horowitz (Chair), Jan de Boer, Ted Jacobson, Mark Van Raamsdonk, Erik Verlinde).

\end{document}